\RequirePackage{ifpdf}
\ifpdf 
\documentclass[pdftex]{sigma}
\else
\documentclass{sigma}
\fi

\usepackage{stmaryrd}
\usepackage[mathscr]{eucal}

\def\BGST{Barnich:2004cr}
\newcommand{\BGadS}{Barnich:2006pc}
\def\BGL{Batalin:2001je}
\def\GL{Grigoriev:2000rn}
\def\Goff{Grigoriev:2006tt}

\newcommand{\dd}{\partial}

\newcommand{\inner}[2]{\langle #1{,}\,#2\rangle}
\newcommand{\binner}[2]{%
  {\langle}\kern-4.15pt{\langle}#1{,}\,#2{\rangle}\kern-4.15pt{\rangle}}

\newcommand{\qcommut}[2]{[#1{,}\,#2]_*}
\newcommand{\pb}[2]{\left\{{}#1{},{}#2{}\right\}}

\newcommand{\half}{\mathchoice{%
    \ffrac{1}{2}}{\frac{1}{2}}{\frac{1}{2}}{\frac{1}{2}}}

\newcommand{\ffrac}[2]{\raisebox{.5pt}%
  {\footnotesize$\displaystyle\frac{#1}{#2}$}\kern1pt}

\newcommand{\brst}{\mathsf{\Omega}}

\newcommand{\derham}{\boldsymbol{d}}

\newcommand{\dl}[1]{\mathchoice{\ffrac{\dd}{\dd #1}}{\frac{\dd}{\dd
      #1}}{\ffrac{\dd}{\dd #1}}{\ffrac{\dd}{\dd #1}}}

\newcommand{\manifold}[1]{\mathscr{#1}}
\newcommand{\manX}{\manifold{X}}

\newcommand{\fR}{\mathbb{R}}
\newcommand{\ZZ}{\mathbb{Z}}

\newcommand{\gh}[1]{\mathrm{gh}(#1)}

\def\vt{\vartheta}
\def\osp{\mathfrak{osp}(2s|2)}
\newcommand{\bundle}[1]{\mathbf{#1}}

\def\cP{\mathcal{P}}

\def\cT{\mathcal{T}}

\begin{document}

\allowdisplaybreaks

\renewcommand{\PaperNumber}{038}

\FirstPageHeading

\ShortArticleName{Manifestly Conformal Descriptions and Higher Symmetries of Bosonic Singletons}

\ArticleName{Manifestly Conformal Descriptions\\ and Higher Symmetries of Bosonic Singletons}

\Author{Xavier BEKAERT~$^\dag$ and Maxim GRIGORIEV~$^\ddag$}

\AuthorNameForHeading{X.~Bekaert and M.~Grigoriev}

\Address{$^\dag$~Laboratoire de Math\'ematiques et Physique Th\'eorique,\\
 \hphantom{$^\dag$}~Unit\'e Mixte de Recherche 6083 du CNRS, F\'ed\'eration Denis Poisson,\\
\hphantom{$^\dag$}~Universit\'e Fran\c cois Rabelais, Parc de Grandmont, 37200 Tours, France}
\EmailD{\href{mailto:xavier.bekaert@lmpt.univ-tours.fr}{xavier.bekaert@lmpt.univ-tours.fr}}

\Address{$^\ddag$~Tamm Theory Department, Lebedev Physics Institute,\\
\hphantom{$^\ddag$}~Leninsky prospect 53, 119991 Moscow, Russia}
\EmailD{\href{mailto:grig@lpi.ru}{grig@lpi.ru}}

\ArticleDates{Received November 09, 2009, in f\/inal form April 23, 2010;  Published online May 07, 2010}

\Abstract{The usual ambient space approach to conformal f\/ields is based on identifying
  the $d$-dimen\-sional conformal space as the Dirac projective hypercone
  in a f\/lat $d+2$-dimensional ambient space.
  In this work, we   explicitly concentrate on singletons of any integer
spin and propose an approach that allows one to have both locality and
conformal symmetry manifest. This is achieved by using the
  ambient space representation in the f\/iber rather than in
  spacetime. This approach allows us to characterize a subalgebra of
  higher symmetries for any bosonic singleton, which is a candidate
  higher-spin algebra for mixed symmetry gauge f\/ields on anti de Sitter spacetime.
  Furthermore, we argue that this algebra actually exhausts all higher symmetries.}


\Classification{70S10; 51P05}

\section{Introduction}

The idea to describe conformal f\/ields in terms of an embedding space
$\fR^{d,2}$ is well known and dates back to
Dirac~\cite{Dirac:1936fq}. In this approach, the conformal
$d$-dimensional space is represented as a quotient of the hypercone
$X^2=0$ modulo the equivalence relation $X\sim \lambda X$ ($\lambda\neq 0$). In this
way, the standard linear action of $\mathfrak{o}(d,2)$ on $\fR^{d,2}$
gives the conformal group action on the projective hypercone.
This
allows one to consider conformal f\/ields as f\/ields def\/ined on
$\fR^{d,2}$ yielding manifestly conformal invariant f\/ield
equations
(see~\cite{Bars:2000mz,Arvidsson:2006fq,Marnelius:2008er,Gover:2009vc}
and references therein for reviews and recent developments).
The conformal space is usually seen as compactif\/ied Minkowski spacetime (as will be done here for concreteness) but it could equally well be identif\/ied with (anti) de Sitter spacetime.

The apparent disadvantage of the embedding approach is the lack of transparent
locality in the sense that the f\/ield theory on the conformal space is formulated in terms
of f\/ields on the ambient space $\fR^{d,2}$. In particular, this leads to
delta-functions in the expressions for f\/ields and Lagrangians.  This
can be partially overcome by considering $\mathfrak{o}(d,2)$ tensor
f\/ields def\/ined in terms of the intrinsic geometry of conformal
space. Such a formulation has been developed~in~\cite{Preitschopf:1998ei} using the conformal analog of the
well-known approach to f\/ield theory on anti de Sitter (AdS) space~\cite{Stelle:1979aj,Preitschopf:1997gi}. In this approach, the
conformal geometry is described in terms of an
$\mathfrak{o}(d,2)$-connection and a conformal compensator f\/ield.
Note also the related developments~\cite{Thomas-BEG} in the context of conformal geometry.

In this work, for a class of conformal f\/ields we develop a formulation such that both conformal invariance and locality are
manifest. It is based on an ambient space construction in the f\/iber
rather than in the spacetime, in the same spirit as Vasiliev's unfolded
description of AdS higher spin
f\/ields~\cite{Vasiliev:2001wa,Vasiliev:2003ev} (for a review, see e.g.~\cite{Bekaert:2005vh}). More precisely, it is the direct
analog of the approach developed in~\cite{\BGadS,\Goff}.  The
geometrical idea underlying such a reformulation is most transparent
in f\/irst-quantized form. In these terms the usual ambient formulation
can be seen as representing the phase space of the conformal particle
as a reduced phase space of a constrained system on the embedding
space.  From this point of view, our approach is based on a
Fedosov-like extension of the constrained system where the embedding
space becomes the f\/iber of a~vector bundle over the true conformal
space.  When formulated in  BRST terms, the theory is determined by a
f\/irst-quantized Becchi--Rouet--Stora--Tyutin (BRST) operator of the same type as the ones describing
Minkowski~\cite{\BGST,AGT} and AdS~\cite{\BGadS,\Goff} 
f\/ields.

In the present work, we restrict ourselves to ``singletons'', i.e.\
f\/ields describing those unitary irreducible representations (UIRs)
of the Poincar\'e group that can be lifted to
UIRs of the conformal group~\cite{Flato:1978qz,Angelopoulos:1997ij,Ferrara:1998bv}. In the case of a singleton with integer spin $s$, the f\/iber constraints originating from the equations of motion and irreducibility conditions form
an orthosymplectic superalgebra $\osp$ and are implemented through the f\/iber BRST operator.
As represented, this $\osp$  superalgebra
is dual to the conformal algebra $\mathfrak{o}(d,2)$ in the sense of Howe duali\-ty~\cite{Howe}. In particular, this guarantees the nilpotency
of the entire BRST operator that consists of the f\/iber part and the f\/lat $\mathfrak{o}(d,2)$-connection. It should be possible to extend the construction to generic conformal (gauge) f\/ields, such as the non-unitary theories of \cite{Fradkin:1985am}, but this is beyond the scope of this paper.

The formalism is then applied to the analysis of the higher
symmetries of spin-$s$ bosonic singletons in dimension $d\geqslant 4$. In
particular we show that the $\mathfrak{o}(d,2)$-traceless invariants of the superalgebra $\mathfrak{osp}(2s|2)$ form a subalgebra of higher
symmetries. Furthermore, for irreducible singletons (i.e.\ those satisfying in addition selfduality conditions) only the invariants belonging to the enveloping of the conformal algebra $\mathfrak{o}(d,2)$ survive.

{\sloppy We provide arguments (but not a mathematically rigorous proof) that $\mathfrak{o}(d,2)$-traceless $\osp$-invariants actually exhaust \textit{all} higher symmetries.
If this turns out to be the case, our result would generalize the theorem of Eastwood~\cite{Eastwood:2002su} to any integer-spin singleton and it should also lead to a generalization of the Flato--Fr\o nsdal theorem \cite{Flato:1978qz,Vasiliev:2004cm} to tensor products of two arbitrary bosonic singletons with identical spin (see~\cite{Boulanger:2008kw} for partial results). Note that in four-dimensional
Minkowski spacetime, all higher symmetries of free massless
helicity-$s$ f\/ields\footnote{For such f\/ields, the algebra of symmetries was identif\/ied in~\cite{Vasiliev:2001zy} with the higher-spin algebra $\mathfrak{hu}_{2s}(1,0|8)\cong\mathfrak{shsc}^0_{2s}(4|0)$ following the respective notations of \cite{Konstein:1989ij,Fradkin:1989xt}. More recently, their conformal and duality symmetries were discussed in~\cite{Vasiliev:2007yc} while
corresponding higher-spin conformal currents were provided in~\cite{Gelfond:2006be,Gelfond:2008ur}.} have been determined for any $s$ in
\cite{Vasiliev:2001zy,Anco:2003ze,Pohjanpelto:2008st}
(see also~\cite{Fushchich:1987rk} and references therein for previous partial results) by using spinorial
techniques.

}

As in the case of the scalar singleton~\cite{Vasiliev:2003ev}, the
algebra of higher symmetries should determine the interactions for AdS$_{d+1}$
gauge f\/ields (so that mixed symmetry f\/ields\footnote{It would be interesting to compare the corresponding spectrum of mixed symmetry f\/ields with the conditions obtained in~\cite{Boulanger:2008kw}.} have to appear in the spectrum corresponding to singleton of spin $s>0$). Another potential application of the formalism has to do with the bulk/boundary correspondence of AdS gauge f\/ields with
conformal operators~\cite{Metsaev:1999ui}.

\section{Conformal f\/ields in terms of the embedding space}

The usual way to make conformal f\/ield equations explicitly conformal
invariant is to def\/ine the f\/ields in terms of the ambient space where
the conformal group acts by linear transformations. In such a
reformulation, the conformal space is constructed as an appropriate
quotient of the hypercone.

More precisely, let $\fR^{d,2}/0$ be the $d+2$-dimensional
pseudo-Euclidean space with the origin excluded. Let $X^A$ ($\,A=-1,0,1,2,\ldots,d-1,d\,$) be standard
coordinates on $\fR^{d,2}$ and $\eta_{AB}={\rm diag}(--++\cdots+)$ the f\/lat
metric. The $d$-dimensional conformal space $\manX$ (conformal
compactif\/ication of the $d$-dimensional Minkowski space) can be
identif\/ied with the subspace $X^2=0$ factorized by the following
equivalence relation $X\sim \lambda  X$ where $\lambda\in \fR/0$ (see e.g.~\cite{Eastwood:2002su} for a review of this
approach). In this way, the standard pseudo-orthogonal algebra
$\mathfrak{o}(d,2)$ in $d+2$ dimensions acts on the conformal space $\manX$ as the
conformal algebra in $d$ dimensions. The action of $\mathfrak{o}(d,2)$
on $\fR^{d,2}$ induces an action on the quotient because both the geometrical
constraint and the equivalence relation are
$\mathfrak{o}(d,2)$-invariant.

\subsection{Conformal scalar f\/ield}\label{cfsc}

To begin with, let us consider the well-known example of the conformal
scalar f\/ield (also called ``scalar singleton''). In order to describe
f\/ields on conformal space, it is convenient to use a f\/irst-quantized
description where f\/ields are identif\/ied with wave functions of the
associated quantum constrained system while equations of motion appear as
constraints (see e.g.~\cite{Barnich:2003wj,\BGST} for a detailed
exposition adapted to the present context). In these terms, the phase
space $T^*\manX$ of a system with conf\/iguration space
$\manX$ can be described in terms of f\/irst-class constraints.  Indeed,
consider the ``ambient'' phase space $T^*(\fR^{d,2}/0)$ with
coordinates $X^A$ and their conjugate momenta $P_B$. The canonical
Poisson bracket reads as $\pb{X^A}{P_B}=\delta^A_B$. The phase space
$T^*\manX$ can be identif\/ied with the reduced phase space of the
constrained system with the following f\/irst-class constraints
\begin{gather}
  X \cdot X=0, \qquad X \cdot P=0 .
\label{constrs}
\end{gather}
Here and in what follows, a ``dot'' will stand for the contraction of
capital Latin indices (by means of the ambient metric if necessary).
For the coordinates $X$, the f\/irst constraint is the geometrical one above (restricting the coordinates $X$ to the light-cone) while the
second one implements the identif\/ication $X\sim \lambda X$ as a gauge
symmetry. For the momenta $P$, the constraint $X\cdot P$ together with the gauge transformation of the momenta $P$ induced by $X^2$
restrict momentum space accordingly,
so that the reduced phase can indeed be identif\/ied with the phase space $T^*\manX$.

At the quantum level, this fact can be seen as follows: suppose the
system is quantized in the coordinate representation, i.e. quantum
states are functions in $X^A$. In Weyl ordering, the constraints
(\ref{constrs}) are implemented as
\begin{gather*}
X^2\Phi=0 , \qquad \left(X \cdot \dl{X} +\frac{d+2}{2} \right)\Phi=0 .
\end{gather*}
The f\/irst constraint implies that physical wave functions can be
represented as $\Phi(X){=}\phi(X) \delta(X^2)$ with $\phi(X)$ def\/ined on the
hypercone $X^2=0$ while the second one takes into account the
equiva\-lence relation above.

In order to describe the scalar particle, one also adds the mass-shell
constraint $P^2$ so that the operators def\/ining the constraints
\begin{gather}
\label{sp2}
 X^2 ,\qquad X \cdot \dl{X}+\frac{d+2}{2} , \qquad \dl{X}\cdot \dl{X}
\end{gather}
form an $\mathfrak{sp}(2)$ algebra. This algebra commutes with the conformal
$\mathfrak{o}(d,2)$ algebra
spanned by the operators $L_{AB}=X_{[A} \dl{X^{B]}}$ where the square bracket stands for an antisymmetrization over a couple of indices, e.g.\ $X_{[A} \dl{X^{B]}}:=X_{A} \dl{X^{B}}-X_{B} \dl{X^{A}}$. This
$\mathfrak{o}(d,2)$ algebra and the $\mathfrak{sp}\,(2)$ algebra form a Howe dual pair~\cite{Howe} on the space
of polynomials in the variables $X^A$.
When imposed on the wave func\-tion~$\Phi(x)$ in some suitable functional space, the constraints (\ref{sp2}) def\/ine the conformal scalar f\/ield
\cite{Marnelius:1978fs} (see
also~\cite{Arvidsson:2006fq,Marnelius:2008er} for more recent
discussions). Note also that this constraint system is extensively
used in the so-called two-time physics \cite{Bars:1998ph} and
in nonlinear higher spin gauge theory~\cite{Vasiliev:2003ev}.

The $\mathfrak{sp}(2)$ covariance can be made even more manifest by using
$\mathfrak{sp}(2)$ indices $m,n=1,2$ to label the coordinates
$W^A_m:=(X^A,P^B)$ of the ``embedding'' phase space
$T^*(\fR^{d,2}/0)\,$. The Poisson bracket then takes the form
$\pb{W_m^A}{W^B_n}=\eta^{AB}\epsilon_{mn}$ with
$\epsilon_{12}=1=\epsilon^{12}\,$. The Weyl symbols for the three
constraints (\ref{sp2}) are the bilinear $\mathfrak{o}(d,2)$-invariants $\eta_{AB}{W_m^A}{W^B_n}$, which span the symmetric
tensor representation of $\mathfrak{sp}(2)$ while the Weyl symbols
for the generators of the conformal algebra are the bilinear $\mathfrak{sp}(2)$
invariants $L^{AB}=X^{[A}P^{B]}=\epsilon^{mn}{W_m^A}{W^B_n}$, which span the antisymmetric
tensor representation of $\mathfrak{o}(d,2)$.

\subsection{Singleton of integer spin}\label{singl}

In this paper, among all conformal f\/ields we are mainly concerned with
singletons. They describe those UIRs of the
Poincar\'e group that can be lifted to UIRs of the conformal group. Here and below the spacetime
dimension $d$ is assumed even (and not smaller than four) because only
scalar and spinor singletons exist in odd dimensions
\cite{Siegel:1988gd,Angelopoulos:1997ij}.

On-shell, the f\/ieldstrength for a
spin-$s$ singleton on Minkowski spacetime takes values in the
f\/inite-dimensional $\mathfrak{o}(d-1,1)$ irreducible representation (irrep) of the Lorentz algebra
$\mathfrak{o}(d-1,1)$ labeled by the rectangular Young diagram made
of $\frac{d}2$ rows of length $s$. As a
Poincar\'e module, the on-shell f\/ieldstrength carries
the UIR of $\mathfrak{iso}(d-1,1)$ associated
to the f\/inite-dimensional
irrep of the isotropy algebra $\mathfrak{o}(d-2)$
labeled by the rectangular Young diagram made of $\frac{d}2-1$ rows
of length $s$. In $d=4$, these f\/ieldstrengths correspond to the
Weinberg--deWit--Freedman curvatures~\cite{Weinberg:1965rz}
so that the four-dimensional spin-$s$ singletons are identical to the massless
helicity-$s$ representations.  In higher dimensions, singletons are very
particular mixed-symmetry gauge f\/ields. Actually, one should further impose the f\/ieldstrength
to be (anti)selfdual in each column in order for the representation to
be irreducible\footnote{This is compatible with a reality condition on the f\/ield only in dimension $d=2$ modulo~4. In any case, we assume f\/ields to be complex valued.}.

In order to describe this f\/ieldstrength in an $\mathfrak{o}(d,2)$-covariant way via the embedding, one should add one more row of length
$s$, as in~\cite{Siegel:1988gd}.  In the antisymmetric basis, the f\/ieldstrength of a singleton of spin $s$ can be described using $s$
anticommuting oscillators carrying the vector $\mathfrak{o}(d,2)$
representation:
\[
\vt_i^A  , \qquad i=1,2,\ldots, s-1,s, \qquad A=-1,0,1,2,\ldots,d-1,d.
\]
Quantum states are functions of $X^A$ and $\vt_i^A$ (therefore they are
``multiforms,'' see e.g.\ the review~\cite{Bekaert:2002jn} on this formalism).

In addition to the operators~\eqref{sp2}, consider the $s^2$ operators
\begin{gather}
 T_i^j=\frac 12 \left(\vt_i\cdot \frac{\partial}{\partial \vt_j}- \frac{\partial}{\partial
\vt_j}\cdot \vt_i\right)=\vt_i\cdot \frac{\partial}{\partial
\vt_j} - \delta_i^j \frac{d+2}{2}
\label{gl}
\end{gather}
that span the $\mathfrak{gl}(s)$ algebra (endowed with the commutator as
Lie bracket). Together with the $s(s-1)$ operators
\begin{gather} \label{oij1}
    T_{ij}=\vt_i\cdot \vt_j ,\qquad
    T^{ij}=\frac{\partial}{\partial \vt_i}\cdot \frac{\partial}{\partial \vt_j}
\end{gather}
the $s(2s-1)$ operators (\ref{gl}) and (\ref{oij1}) span the
$\mathfrak{o}(2s)$ algebra (endowed with the commutator
as Lie bracket). This algebra $\mathfrak{o}(2s)$ commutes with the conformal
algebra $\mathfrak{o}(d,2)$ spanned by the operators
$S_{AB}=\vt_{i [A} \dl{\vt^{B]}_i}$
where there is a sum corresponding to the repeated index~$i$.  This
$\mathfrak{o}(d,2)$ algebra and the $\mathfrak{o}(2s)$ algebra
form a Howe dual pair on the space
of polynomials in the fermionic variables $\vt_i^A$.  In this space,
the kernel of the operators (\ref{gl}) and (\ref{oij1}) def\/ines a
f\/inite-dimensional representation of the algebra $\mathfrak{o}(d,2)$
labeled by the traceless rectangular Young diagram made of $\frac{d}{2}+1$ rows
of length~$s$.

Adding the $4s$ operators
\begin{gather}\label{xvt}
X\cdot \vt_i
    ,\qquad
X\cdot \frac{\partial}{\partial \vt_i}
 ,\qquad
\frac{\partial}{\partial X}\cdot \vt_i,\qquad
\frac{\partial}{\partial X}\cdot \frac{\partial}{\partial \vt_i},
\end{gather}
the set of $2s^2+3s+3$ operators \eqref{sp2}--\eqref{xvt}
span the $\mathfrak{osp}(2s|2)$ superalgebra endowed with the graded
commutator as bracket.
This algebra commutes with the conformal $\mathfrak{o}(d,2)$ algebra
spanned by the operators
\begin{gather}
J_{AB}=L_{AB}+S_{AB}=X_{[A}\dl{X^{B]}}+ \vt_{i[A} \dl{\vt^{B]}_i},
\label{conf}
\end{gather}
where there is a sum corresponding to the repeated index $i\,$.
More precisely, the conformal al\-geb\-ra $\mathfrak{o}(d,2)$ and the
$\mathfrak{osp}(2s|2)$ superalgebra form a Howe dual pair
on the space of polynomials in~$X^A$ and~$\vt_i^A$.
In multiform language, the operators ${\partial}/{\partial X}\cdot \vt_i$
(and ${\partial}/{\partial X}\cdot {\partial}/{\partial \vt_i}$) are
(respectively) identif\/ied with the de Rham (co)dif\/ferential in the $i$-th
column.
Their kernel def\/ine harmonic multiforms, as in the generalized
Bargmann--Wigner equations of~\cite{Bekaert:2002dt,Bandos:2005mb}.

In some proper class of functions in $X$ and $\vt$,
the kernel of the operators (\ref{sp2})--(\ref{xvt}) carries a~spin-$s$
singleton representation in $d$ dimensions~\cite{Arvidsson:2006fq}. These f\/ield equations can be interpreted as the conformal covariantization (along similar lines to the
seminal work \cite{Siegel:1988gd}) of the
Poincar\'e covariant equations \cite{Bekaert:2002dt,Bandos:2005mb}.
As far as general conformally invariant wave equations are concerned, a~rather complete classif\/ication
using unfolding has been provided in \cite{Shaynkman:2004vu}.
The orthosymplectic superalgebra $\mathfrak{osp}(2s|2)$  f\/inds a natural interpretation in the
$\mathfrak{o}(2s)$ extended supersymmetric spinning particle (see e.g.~\cite{Burkart:2008bq,Bastianelli:2009eh,Marnelius:2009uw}
for recent developments). In dimension $d=4$, the quantum constraints appearing in this model give rise to proper f\/ield equations for the massless helicity-$s$ particles~\cite{Gershun:1979fb}.

Again, the underlying algebras can be made more manifest by making use of an extension of the $\mathfrak{sp}(2)$ indices.
Quantum states are functions on the
$d+2|s(d+2)$ dimensional superspace with $d+2$ even coordinates $X^A$
and $s(d+2)$ odd ones $\vt_i^A$.
Let $(P_A|\pi^i_B)$ be the conjugates of the supercoordinates $(X^A|\theta^B_i)$.
The phase (super)space coordinates are collectively denoted by $Z^A_{\alpha}:=(X^A,P_B|\theta_i^A,\pi^j_B)$ where
the superindices $\alpha$, $\beta$ take $2+2s$ values.
The indices~$A$,~$B$ correspond to a $(d+2)$-dimensional
pseudo-Euclidean space ${\bf V}$ with metric $\eta^{AB}$,
while the superindices $\alpha$, $\beta$ correspond to a $(2|2s)$-dimensional symplectic superspace
${\bf U}$ with symplectic form $\epsilon_{\alpha\beta}$.
The phase space coordinates $Z^A_\alpha$ are natural coordinates on the tensor
product ${\bf V}\otimes {\bf U}$. The graded Poisson bracket originating from the symplectic
structure on the phase superspace is given by
$\big\{Z_\alpha^A,Z^B_\beta\big\}=\eta^{AB}\epsilon_{\alpha \beta}$.
The metric form on ${\bf V}$ is preserved by the algebra $\mathfrak{o}(d,2)$ while
the symplectic form $\epsilon_{\alpha \beta}$ on the $(2|2s)$-dimensional superspace ${\bf U}$
can be seen as a metric structure on the $(2s|2)$-dimensional superspace $\Pi {\bf U}$
(with opposite Grassmann parity), so it is preserved by the orthosymplectic algebra
$\mathfrak{osp}(2s|2)$. The following bilinears
\begin{gather*}
t_{\alpha\beta}=\eta_{AB}Z^A_{\alpha}Z^B_{\beta},\qquad
J^{AB}=\epsilon^{\alpha \beta}Z^A_{\alpha}Z^B_{\beta},
\end{gather*}
are respectively the generators of $\mathfrak{osp}(2s|2)$ and $\mathfrak{o}(d,2)$.  In fact, these can be also seen as Weyl
symbols of the operators (\ref{sp2})--(\ref{xvt}) and \eqref{conf}
respectively.

\section{Formulation in terms of intrinsic geometry}

\subsection{The general method: ambient space as f\/iber}

In order to formulate the theory in terms of f\/ields explicitly def\/ined
on the conformal space with local coordinates $x^\mu$ ($\mu=0,1,\ldots,d-1$) we use a similar
procedure to~\cite{\BGadS}. We replace the spacetime variables
$X^A$ with $Y^A+V^A$, where $Y^A$ are formal variables and $V^A(x)$ is
a light-like vector f\/ield: the ``conformal compensator''\footnote{The
conformal compensator has been introduced in~\cite{Preitschopf:1998ei} in the context of conformal gravity.}
satisfying $V^2=0$.

More precisely, consider the vector bundle $\bundle{V}(\manX)$ over the conformal space $\manX$ with the f\/iber being a $(d+2)$-dimensional pseudo-Euclidean vector space equipped with the bilinear form $\eta$. The variables $Y^A$ are then identif\/ied
as coordinates on the f\/ibers with respect to a given local frame $E_A$ while
$V^A(x)$ are seen as components of a given section of $\bundle{V}(\manX)$.
In addition, $\bundle{V}(\manX)$ is equipped with a f\/lat $\mathfrak{o}(d,2)$-connection
$\omega$ and the vielbein $e$ with coef\/f\/icients given by~$\omega_{\mu A}^B$ and~$e_\mu^A$. These satisfy:
\begin{alignat}{3}
& \omega \eta + \eta  \omega^T =0 ,\qquad  && \derham\omega+\omega^2=0 , & \nonumber\\
& \derham e+\omega e =0 ,\qquad &&e=\derham V+\omega V. & \label{eq:flat}
\end{alignat}
In addition, $e_\mu^A$ is required to be of maximal rank (i.e.\ ${\rm rank}(e_\mu^A)=d$).
The last equation expresses the vielbein through $\omega$ and $V$. The connection $\omega$ and the compensator $V$ can be seen as def\/ining conformal structure on $\manX$ in the sense of Cartan. Note that the null vector f\/ield $V$ is naturally def\/ined up to a rescaling by a nonvanishing factor. This originates from the identif\/ication of $V^A$ with the Cartesian coordinates $X^A$ on the ambient space (seen as functions on a submanifold of the hypercone). The arbitrariness ref\/lects the ``gauge'' freedom $X \to \lambda X$.

One next considers an associated vector bundle with the f\/iber
being the space of formal power series in
$Y$ variables tensored with the Grassmann algebra of $\vt$
variables.  In the f\/iber the covariant derivative acts as follows
\begin{gather}
 \nabla=\derham-
dx^\mu e_\mu^A\dl{Y^A}- {dx^\mu} \omega_{\mu A}^B(x)\Big(Y^A\dl{Y^B}+\vt^A_i\dl{\vt^B_i}\Big)
\nonumber \\
\phantom{\nabla}{} =\derham-
dx^\mu \omega_{\mu A}^B(x)\Big(\big(\,Y^A+V^A\,\big)\dl{Y^B}+\vt^A_i\dl{\vt^B_i}\Big),
 \label{covder}
\end{gather}
where in the second equality we made use of~\eqref{eq:flat} and assumed that the local frame is chosen such that
$V={\rm const}$.

As a next step one replaces the original constraints \eqref{sp2}--\eqref{xvt} with the corresponding constraints in the f\/iber. This
is achieved by replacing, respectively, $X$ with $Y+V$ and $\partial/\partial X$ with $\partial/\partial Y$ in the
expression for the constraints. Notice that the covariant derivative commutes with the
constraints in the f\/iber because $\mathfrak{o}(d,2)$ and $\mathfrak{osp}(2s|2)$ are Howe dual
on this representation space as well. Such a reformulation has been suggested
in \cite{\BGadS,\BGST} in the context of f\/ields on
constant-curvature space. From the f\/irst-quantized point of view,
it is {somewhat} analogous to the reformulation used in the Fedosov
quantization~\cite{Fedosov-book} (see also~\cite{\GL,\BGL} for the
generalization of the Fedosov approach to the case of systems with
constraints). At the  same time the proposed approach is related to the tractor bundle
approach. In particular, the bundle $\bundle{V}(\manX)$ introduced above is known
in the context of conformal geometry as the tractor bundle. However, in
contrast to the tractor
approach \cite{Thomas-BEG,Gover:2008sw,Gover:2009vc}, sections of the associated bundle considered above are not tractor
tensors because of the the unusual (involving $V^A$) transformation law.

More technically, one now introduces the BRST operator
\begin{gather*}
 \brst=\nabla + \bar\brst,
\end{gather*}
where $\nabla$ is the covariant derivative (\ref{covder}) with the
basis one-forms $dx^\mu$ replaced by the fermionic ghost variables
$\theta^\mu$ with $\gh{\theta^\mu}=1$ and $\bar\brst$ is a
BRST operator implementing the constraints in the f\/iber. It is
assumed that the f\/iber is extended to incorporate the ghost
variables associated to the f\/iber constraints. Note that the BRST
operator $\brst$ is automatically nilpotent because~$\nabla$ and~$\bar\brst$ are respectively the f\/lat connection and the BRST
operator of two commuting Lie algebras (in this case Howe dual
algebras). It is important to stress that it often appears useful
to impose a subset of the constraints (strictly speaking their
BRST-invariant extension) directly on states and hence not to
incorporate them into $\bar\brst$.

The physical f\/ields of the model are contained in the ghost number zero wave function $\Psi^{(0)}$ while the equations of motion and gauge symmetries read as
\begin{gather*}
 \brst \Psi^{(0)}=0 ,\qquad \delta_\Lambda \Psi^{(0)} =\brst \Lambda,
\end{gather*}
where the gauge parameter $\Lambda$ is identif\/ied with a general state of ghost number $-1$.  Analogously, $\brst$ determines the higher order reducibility relations
for gauge symmetries and equations of motion (see e.g.~\cite{\BGST} for more details
and precise def\/initions).

The procedure described above gives an elegant way to formulate
the conformal f\/ields in manifestly local terms, using f\/ields
def\/ined on the conformal space. This procedure is rather general and allows to
consider f\/ields on Minkowski and AdS space as well as conformal
f\/ields. Here we explicitly show how it works for the particular
class of conformal f\/ields described above: bosonic singletons. To begin with we
explicitly consider the simplest but typical example of the
conformal scalar f\/ield.

\subsection{An example: scalar singleton}

The constraints \eqref{sp2} become the following constraints in the f\/iber:
\begin{gather}\label{sp2y}
\overset{{}_-}{\Box} =(Y+V)\cdot(Y+V),\qquad h=(Y+V)\cdot \dl{Y}+\frac{d+2}{2},\qquad
\Box = \dl{Y}\cdot\dl{Y}.
\end{gather}
Introducing Grassmann odd ghost variables $c_+$, $c_0$, $c_-$ with $\gh{c_+}=\gh{c_0}=\gh{c_-}=1$ and their conjugate ghost momenta $b_+$, $b_0$, $b_-$ let us represent them on polynomials in $c_+$, $c_0$, $b_-$ according to
\begin{gather}
\label{bbcc}
 b_{+}=\dl{c_+},\qquad b_0=\dl{c_0},\qquad c_-=\dl{b_-}.
\end{gather}
Constraint $\overset{{}_-}{\Box}$ cannot be imposed directly on states
because its kernel is empty. Therefore it
has to be quotiented out rather than
directly imposed. In BRST terms, this is implemented by
representing the ghost variable~$c_-$ in the momentum
representation, as in~\eqref{bbcc}.

The BRST operator takes the form
\begin{gather}
\brst=\derham-e^A\dl{Y^A}-\omega^A_B Y^B\dl{Y^A}
+c_+\Box+c_0(h-2)+\overset{{}_-}{\Box}\dl{b_-}+\text{ghosts}.
\label{brstomega}
\end{gather}
Here ``ghosts'' stand for the purely ghost terms originating from the
structure constants of~$\mathfrak{sp}(2)$ and ordered such that they
annihilate constants. The representation space where $\brst$ acts is the space of functions
in~$x$,~$\theta$ with values in the formal series in $Y$-variables tensored with polynomials in ghosts $c_+$,~$c_0$,~$b_-$.
Note that the ordering constant in the coef\/f\/icient of~$c_0$ is uniquely f\/ixed by the nilpotency of the BRST charge
once the representation on which it acts is f\/ixed.

A natural way to see which dynamics is described by the BRST
model (\ref{brstomega}) is to reduce it to the standard Lorentz covariant formulation. To this end let us f\/irst construct a useful local frame for the vector bundle $\bundle{V}(\manX)$. Let $E_+=V$ and $E_-$
be two sections such that $\eta(E_+,E_-)=1$ and $\eta(E_+,E_+)=\eta(E_-,E_-)=0$ at a given point $p$. One can also assume $\nabla E_-=0$ and hence $\eta(E_-,E_-)=0$ everywhere. Indeed, $E_-$ can be extended to a neigbourhood of $p$ by a covariant constancy condition.
By rescaling $V$, one can then achieve $\eta(E_+,E_-)=1$ in a neigbourhood of~$p$. Note that $V^2=0$ implies $\eta(E_+,E_+)=0$.
One takes the remaining
local sections $E_a$ to be orthogonal to $E_+$ and $E_-$ and such that
$\eta(E_a,E_b)={\rm diag}(-++\cdots+)$. In the constructed frame, the components of $V$ and $\eta$ read as
\begin{gather}
\begin{split}
& V^+=1,\qquad V^-=0,\qquad V^a=0, \\
& \eta_{+-}=1,\qquad
\eta_{++}=\eta_{--}=\eta_{a+}=\eta_{a-}=0,\qquad
\eta_{ab}={\rm diag}(-++\cdots+),
\end{split}\label{frame}
\end{gather}
i.e.~$E_+$, $E_-$, $E_a$ form a light-cone-like basis at each point.
As we are going to see $a,b=0,1,\ldots,d-1$ are to be seen as Lorentz indices.

Introducing the connection components $\omega^A_B=dx^\mu \omega_{\mu B}^A$
(where $A:=(+,-,a)\,$) with respect to the frame $E_\pm$, $E_a$, one observes
that $\omega_-^a=\omega_-^\pm=0$ because $\omega_-^A=(\nabla E_-)^A$ by def\/inition
and $\nabla E_-=0$ by construction.
Together with
\eqref{eq:flat} this imply
\begin{gather*}
\omega_+^a=-\eta^{ab}\omega^-_b=e^a,\qquad \omega_-^a=\omega^+_a=\omega^+_+=\omega_-^-=0=e^\pm,\nonumber\\
\omega_a^c \eta^{}_{cb}+\eta^{}_{bc}\omega^c_a=0,\qquad
d\omega^a_b+\omega^a_c\omega^c_b=0,
\qquad de^a+\omega^a_b e^b=0.
\end{gather*}
Therefore $\manX$ can locally be seen as the
Minkowski space\footnote{The conformal space $\manX$ can also be seen as
(A)dS space. If the timelike component $V^{-1}$ is equal to a constant, say $R$,
then the other components $V^{\hat A}$ ($\hat A,\hat B=0,1,2,\ldots,d-1,d$)
of the light-like conformal compensator $V^A$ such that $\eta_{AB}V^A V^B=0$
can be interpreted as the components of the space-like de Sitter
compensator obeying $\eta_{\hat A\hat B}V^{\hat A} V^{\hat B}=(V^{-1})^2=R^2$.
For AdS space of curvature radius $R$, one should consider instead $V^d=R$ as constant.
One indeed f\/inds the usual description of (anti) de Sitter geometry in terms
of the appropriate f\/lat connection and the compensator f\/ield.}
provided
one identif\/ies $e_\mu^a$, $\omega_{\mu a}^b$ as the coef\/f\/icients of
the vielbein and the Lorentz connection.

Using the Lorentz adapted frame \eqref{frame} allows one to reduce
the system to the standard Lorentz covariant formulation.
To this end, one applies the method of homological
reductions~\cite{\BGST,\BGadS} to eliminate the auxiliary f\/ields
and the St\"uckelberg f\/ields present in the model. The reduction is
done in two steps.

As a f\/irst step, let us reduce the theory to the
cohomology of the nilpotent operator
\begin{gather*}
\brst_{-1}:=\overset{{}_-}{\Box}\dl{b_-}=(Y\cdot Y +2V\cdot Y)\dl{b_-}
\end{gather*}
entering the BRST operator $\brst$. This is the term of lowest degree $(-1)$
with respect to the degree ``$\deg$'' def\/ined by $\deg{b_-}=\deg{c_+}=1$ (and
degree zero for all other variables).
In the space of formal power series, one can show that
representatives of $\brst_{-1}$-cohomology classes are~$b_-$
independent and can be chosen $Y^-$-independent. In order to f\/ind
the BRST operator of the reduced system one notices that
$\brst$ does not map such representatives into the
image of~$\brst_{-1}$. One then concludes that the reduced
operator is just the restriction of $\brst$ to the space of
$b_-$, $Y^-$-independent elements and is given explicitly by
\begin{gather*}
  \brst^{\text{1st step}}=\derham-\omega^a_b Y^b\dl{Y^a}-
\theta^\mu e_\mu^a (Y^+ +1)\dl{Y^a} \nonumber
\\
\phantom{\brst^{\text{1st step}}=}{} + c_+\Box_0 +
c_0\Big(Y^a\dl{Y^a}+(Y^++1)\dl{Y^+}+\ffrac{d-2}{2}\Big)+2c_0
c_+\dl{c_+},
\end{gather*}
where $\Box_0=\dl{Y^a}\dl{Y_a}$.

As a second step, we reduce the model to the cohomology of the term
$\brst^{\text{1st step}}_{-1}=c_0\dl{Y^+}$. It can be seen as the degree $-1$
term in $\brst^{\text{1st step}}$ when the grading is the homogeneity degree in the variable $Y^+$ (such that
${Y^+}$ carries degree $1$ and all other variables vanishing degree).
Its cohomology is
given by $c_0$, $Y^+$-independent elements and the reduced system is
described by
\begin{gather*}
  \brst^{\text{2nd step}}=\derham-\omega^a_b Y^b\dl{Y^a}-
\theta^\mu e_\mu^a \dl{Y^a}+
c_+ \Box_0  .
\end{gather*}
It was shown in~\cite{\BGST} that this BRST operator is that of the parent
theory constructed for a~BRST f\/irst-quantized system with the BRST operator
\begin{gather*}
{\brst}^{\text{standard}}=c_+\nabla^\mu\nabla_\mu,
\end{gather*}
where $\nabla_\mu$ denotes the standard f\/lat and torsion-free
covariant derivative determined by $\eta$, $e$, $\omega$. This shows
that the model indeed describes a massless scalar f\/ield on
Minkowski space.

\subsection{General case: singleton of integer spin}

The case of arbitrary ``spin'' $s$ (more precisely, $s$ is the
number of columns in the Young diagram labelling the corresponding
irrep) is completely analogous to the case $s=0$ above but is more
involved technically. In addition to the constraints~\eqref{sp2y}
one has constraints \eqref{gl}, \eqref{oij1} which are unchanged
and constraints \eqref{xvt} taken to the f\/iber:
\begin{gather*}
\bar S_i =(Y+V) \cdot \vt_i, \qquad
\bar S^i=(Y+V)\cdot \dl{\vt_i},
\qquad
S_i=\vt_i\dl{Y},
 \qquad
S^i=\dl{\vt_i}\cdot \dl{Y},
\end{gather*}
where we have also introduced some notations.

These constraints are to be consistently imposed in the f\/iber. This is achieved
through a~BRST operator of the form
\begin{gather}
\label{barbrst}
\bar\brst= C^It_I+\half U^K_{IJ}C^IC^J \cP_K,
\end{gather}
where the generators $t_I$ stand for all the constraints
$\Box$, $\overset{{}_-}{\Box}$, $h$, $S$, $\bar S$, $T$ of $\mathfrak{osp}(2s|2)$ while
$U^K_{IJ}$ stand for its structure constants and $C^I$, $\cP_I$ for the conjugated
ghost variables. If ghosts are represented as $\cP_I=\dl{C^I}$ on
polynomials in $C^I$ this BRST operator is simply the standard Lie
(super)algebra dif\/ferential, with the representation space being functions
in $Y^A$, $\vt^A_i$.

As in the case of the conformal scalar f\/ield, constraints
$\overset{{}_-}{\Box}$ and $\bar S_i$ are to be quotiented out rather than
directly imposed. In BRST language, this is implemented by
representing the respective ghost variables in the momentum
representation. Furthermore, all the constraints not involving
$Y$-variables are ``of\/f-shell constraints'' in the sense that they do
not lead to dynamical equations. If such constraints are
present this can give rise to doubling of physical states due to
additional cohomology classes\footnote{A well known example is
the level-matching constraint in closed string f\/ield theory.}.
To be on the safe side, in addition to the BRST invariance one
needs to require physical states not to depend on the ghost
variables associated to the of\/f-shell constraints. Equivalently,
one can impose such constraints  (strictly speaking their BRST-invariant extensions) directly on states. In the present case the
latter option appears more instructive and compact.

In the case at hand, the of\/f-shell constraints are
$T_{ij}$, $T^{ij}$, $T^i_j$ given by~\eqref{gl},~\eqref{oij1}. The remaining
constraints $\Box$, $h$, $\overset{{}_-}{\Box}$ and $S_i$, $\bar S^i$ are introduced
through the following reduced BRST operator
\begin{gather}
\label{brst-g}
 \hat\brst= c_+ \Box + c_0 (h-2+s)+\overset{{}_-}{\Box} \dl{b_-}
+\bar\gamma_i S^i+\bar \gamma^i S_i +  \gamma_i \bar S^i+\bar
S_i\dl{\beta_i}+\mbox{ghosts},
\end{gather}
where, again, ``ghosts'' denotes the pure ghost terms originating
from the structure constants. The representation space is the
subspace of functions in $Y$, $\vt_i$, and ghost variables
$c_+$, $c_0$, $b_-$, $\bar\gamma_i$, $\bar\gamma^i$, $\gamma_i$, $\beta_i$ singled
out by the $\hat\Omega$-invariant extensions of the constraints
$T_{ij}$, $T^{ij}$, $T^i_j$. Note also that ghost variables
$c_+$, $c_0$, $b_-$ are fermionic while
$\bar\gamma_i$, $\bar\gamma^i$, $\gamma_i$, $\beta_i$ are bosonic. The
ghost degrees are as follows:
\begin{gather}
\label{ghosts}
\gh{c_+}=\gh{c_0}=\gh{\gamma}=\gh{\bar\gamma}=1,\qquad
\gh{b_-}=\gh{\beta_i}=-1.
\end{gather}
In more details, the extended of\/f-shell constraints read as
\begin{gather}
\cT_{ij} :=
\vt_i\cdot\vt_j+\bar\gamma_i\dl{\bar\gamma^j}
-\bar\gamma_j\dl{\bar\gamma^i}-\gamma_i \beta_j+\gamma_j\beta_i ,\nonumber\\
\cT^{ij} :=
\dl{\vt_i}\cdot\dl{\vt_j}+\bar\gamma^i\dl{\bar\gamma_j}
-\bar\gamma^j\dl{\bar\gamma_i}
+\dl{\beta_i}\dl{\gamma_j}-\dl{\beta_j}\dl{\gamma_i} ,  \label{cT12}
\end{gather}
and
\begin{gather}
\label{cT3}
 \cT^i_j:=
\vt_i\cdot\dl{\vt_j}-\frac{d}{2}  \delta_i^j
+\bar\gamma_i\dl{\bar\gamma_j}
-\bar\gamma^j\dl{\bar\gamma^i}
-\beta_i\dl{\beta_j}
-\gamma_i\dl{\gamma_j} .
\end{gather}

The following comments are in order:
$(i)$~The $\hat\brst$-invariant extensions $\cT$ form the same algebra
as the nonextended constraints $T$. $(ii)$~The operator $\hat\Omega$ is nilpotent only in the subspace of elements annihilated by
the constraints $\cT$, because these constraints appear in the commutators of~$S$ and~$\bar S$.
$(iii)$~It is assumed that the term ``ghosts''  in \eqref{brst-g}, representing the pure ghost terms, is ordered such that it annihilates constants. $(iv)$~The shift in the constant term in $\cT^i_i$ as well as the constant correction to~$h$ in~$\hat\brst$  originate from the change of the representation for ghosts associated to $\bar S_i$ and~$\bar \Box$.

As was discussed above, another way to arrive at \eqref{brst-g} and constraints \eqref{cT12} and \eqref{cT3}
is to start with the total BRST operator \eqref{barbrst} and decompose it as
\begin{gather*}
 \bar\brst=\hat\brst+\brst_\cT+\text{``extra''},
\end{gather*}
where $\brst_\cT$ denotes the usual BRST operator implementing constraints $\cT$
using their own ghost variables $\xi$ and ``extra'' denotes the terms involving
$\dl{\xi}$ and the ghosts {from} \eqref{ghosts} (these terms ref\/lect the fact that constraints $\cT$ appear in the commutators of $S$ and $\bar S$). It is clear that
$\hat\brst$ acts in the subspace of f\/ields $\phi$ such that $\cT\phi=0=\dl{\xi}\phi$ and it is nilpotent
there.

As we are going to see, the equations of motion determined by
BRST operator $\hat\brst$ and constraints $\cT$ (and hence by BRST operator \eqref{barbrst}) indeed
describe the correct degrees of freedom. As in the example of
the conformal scalar f\/ield, it is useful to reduce the system to the equivalent form adapted to Minkowski space. Also the reduction does not really af\/fect the term $\nabla$ in the BRST dif\/ferential $\brst$ (it is passive in the reduction)
so that one can simply concentrate on reducing the part $\hat\brst$.
In what follows, we again work in the
light-cone frame~\eqref{frame}.

As a f\/irst step of the reduction, let us consider the following degree
\begin{gather*}
 \deg{b_-}=\deg{c_+}=\deg{\beta_i}=\deg{\bar\gamma_i}=1.
\end{gather*}
The lowest degree component of the BRST operator is then
\begin{gather*}
 \hat\brst_{-1}=(Y+V)^2\dl{b_-}+(Y+V)\cdot\vt_i\dl{\beta_i}.
\end{gather*}
The standard arguments of homological perturbation theory (see e.g.~\cite{Barnich:2004cr,\Goff} for more details on analogous cases) then show that the representatives of the
{$\hat\brst_{-1}$}-cohomology can be chosen~$b_-$,~$\beta_i$ and
$Y^-$, $\vt^-_i$-independent. The reduced theory is then determined by the
BRST operator~$\hat\brst^{\text{1st step}}$ (that is simply $\hat\brst$
reduced to the $b_-$, $\beta_i$ and $Y^-$, $\vt^-_i$-independent subspace).

As a second step one chooses a grading such that
$Y^+$ and $\vt^+_i$ carry degree $1$. The lowest degree term is then
\begin{gather*}
\hat\brst^{\text{1st step}}_{-1}=c_0\dl{Y^+}+\gamma_i\dl{\vt^+_i}.
\end{gather*}
The standard arguments show that its cohomology representatives can be chosen not to depend on the variables $c_0$, $Y^+$, $\gamma_i$, $\vt^+_i$. In terms of such representatives the reduced system is deter\-mi\-ned by the following reduced BRST operator:
\begin{gather*}
 \hat\brst^{\text{2nd step}}=c_+\Box_0 +\bar\gamma^i
S^0_i+\bar\gamma_i S_0^i-\bar\gamma^i\bar\gamma_i\dl{c_+}\nonumber\\
 \phantom{\hat\brst^{\text{2nd step}}}{}=
 c_+\dl{Y^a}\dl{Y_a} +\bar\gamma^i
 \vt^a_i\dl{Y^a}
 +\bar\gamma_i \dl{Y^a}\dl{\vt^a_i}-\bar\gamma^i\bar\gamma_i\dl{c_+}.
\end{gather*}

As above, in order to see the explicit form of the equations of
motion in terms of the usual Cartesian coordinates $x^a$ on the Minkowski
space, one f\/irst observes that once variables
$Y^+$, $Y^-$, $\vt^+_i$, $\vt^-_i$ are eliminated the covariant
dif\/ferential (\ref{covder}) reduces to the usual Minkowski space one. In this way
the reduction of the total BRST operator reads as{\samepage
\begin{gather}
\label{reduced}
 \brst^{\text{reduced}}=
\theta^{a}\left(\dl{x^a}-\dl{Y^a}\right)
+\hat\brst^{\text{2nd step}},
\end{gather}
since $e_a^b=\delta_a^b$ and $\omega_a^b=0$.}

One then eliminates variables $Y^a$, $\theta^a$ using the reduction
described in~\cite{\BGST}. At the  end, this simply amounts to
replacing $\dl{Y^a}$ with $\dl{x^a}$ and putting to zero
$\theta^a$ and~$Y^a$.  For the ghost number zero element $\phi$, one then
gets the following equations of motion
\begin{gather*}
 \dl{x^a}\dl{x_a} \phi=0,\qquad \dl{\vt_i^a}\dl{x_a} \phi=0,\qquad
\vt_i^a\dl{x^a} \phi=0.
\end{gather*}
For a ghost number zero f\/ield, of\/f-shell constraints $\cT$ coincide with constraints $T$
up to an ordering constant in $\cT^i_j$. They explicitly give
\begin{gather}
 \dl{\vt^a_i}\dl{\vt_{j\,a}}\phi=0,\qquad \vt^a_{i} \vt_{j\,a} \phi=0,\qquad
\vt^a_i \dl{\vt^a_j}\phi=0,\qquad
\left(\vt^a_i\cdot \dl{\vt^a_i}-\frac{d}{2} \right)\phi=0.\label{g-const}
\end{gather}
These conditions say that $\phi(x,\theta)$ is a singlet of
the orthogonal algebra $\mathfrak{o}(2s)$ formed by the operators
$T_{ij}$, $T^{ij}$ and $T^i_j+\delta^i_j$ def\/ined only in terms of the $d$-dimensional fermionic oscillators
$\vt^a_i$. Thus (\ref{g-const}) implies that we are dealing with tensor f\/ields
represented by $s\times \frac{d}{2}$ rectangular Young tableaux.
These f\/ields can also be seen as curvatures of the respective
gauge f\/ields~\cite{Bekaert:2002dt,Bandos:2005mb,Arvidsson:2006fq}.

\subsection{Gauge description for multiforms}

The above formulation is in general not Lagrangian. This is because the physical f\/ields are gauge invariant curvatures rather than gauge potentials.
The reformulation in terms of gauge potentials in general breaks conformal invariance~\cite{Bracken:1982ny} (see~\cite{Vasiliev:2007yc} for the conformally invariant description in ${\rm AdS}_4$ though).

In terms of the reduced theory, the gauge description can be achieved
as follows. Let us change the representation for ghost variables $\bar\gamma^i$
according to $\bar\gamma^i \to \dl{\bar\beta_i}$, $\dl{\bar\gamma^i} \to -\bar\beta_i$ and keep the ghost degree prescription in the representation space
(i.e.\ the ghost number of a state described by a ghost-independent function vanishes).
This brings the BRST operator \eqref{reduced} to the following form\footnote{This is a direct analog of the BRST operator~\cite{AGT} describing mixed-symmetry f\/ields on Minkowski space in terms of bosonic oscillators.}
\begin{gather}
\label{brst-gauge}
\tilde\brst=c_+\dl{x^a}\dl{x_a} +\vt^a_i\dl{x^a}\dl{\bar\beta_i}+\bar\gamma_i\dl{x_a}\dl{\vt^a_i}-\bar\gamma_i\dl{\bar\beta_i}\dl{c_+}.
\end{gather}
where we have explicitly eliminated
variables $\theta^a$, $Y^a$. One should recall that states are assumed to be annihilated
by the constraints $\cT$. Note that, actually,
the constraints $\cT^i_i$ get additional constant
contributions originating from the change of representation for ghosts.

One observes that $\tilde\brst$ can be
made formally hermitian with respect to the natural inner pro\-duct.
This implies that this system is in fact Lagrangian with an
action of the form $S=\half\inner{\Psi}{\tilde\brst\Psi}$,
where $\Psi$ belongs to the subspace annihilated by the $\tilde\brst$-invariant version of constraints~$\cT$. Let us stress that one can replace  constraints~$\cT$
with more general irreducibility conditions in order to describe arbitrary mixed-symmetry tensor f\/ields on Minkowski space in the same way as in~\cite{AGT}. The easiest way to see that this is indeed {possible} is to reduce the theory determined by~$\tilde\brst$ to a~light-cone gauge. The reduction is algebraically completely analogous to the light-cone reduction performed in~\cite{AGT}  in the case of
mixed-symmetry f\/ields described in terms of bosonic oscillators (see also~\cite{Barnich:2005ga} for a more detailed discussion in the case of totally symmetric f\/ields).

Let us stress that the action $S=\half\inner{\Psi}{\tilde\brst\Psi}$ with $\tilde\brst$ given by \eqref{brst-gauge}
and $\Psi$ satisfying the BRST invariant extensions of the constraints \eqref{g-const} is \textit{not} conformally invariant for $s>1$. This is just a usual action for a particular mixed symmetry f\/ields on Minkowski space, actions which are known to be not conformal for a number of columns $s>1$. Indeed, in this case the respective equations of motion do not f\/it into the exhaustive classif\/ication of the free conformal equations obtained in~\cite{Shaynkman:2004vu} (see also~\cite{Bracken:1982ny} for an early work on the $d=4$ particular case).

In the case $s=1$, the only of\/f-shell constraint is given by
${\cal T}=\vt^a\dl{\vt^a}-\frac{d-2}{2}+\bar\beta\dl{\bar\beta}+\bar\gamma\dl{\bar\gamma}$
so that indeed the BRST operator \eqref{brst-gauge} determines the Lagrangian gauge theory of
$(\frac{d}{2}-1)$-form f\/ield known in the litterature. This formulation was analyzed in details
in~\cite{Buchbinder:2008kw} (see also \cite{Henneaux:1987cpbis}) for the case of general $p$-form f\/ields,
so that we skip the details\footnote{In order to help the reader interested in this explicit procedure,  Section~2 of \cite{Buchbinder:2008kw} addresses precisely this issue and one can easily make contact with their notations through the following translation rules: ${\vt^\mu}\rightarrow a^{+\mu}$, $\dl{\vt^a}\rightarrow a^\mu$, $c_+\rightarrow \eta_0$, $\dl{c_+}\rightarrow {\cal P}_0$, $\bar\gamma_1\rightarrow -iq^+_1$,  $\dl{\bar\gamma_1}\rightarrow -p_1$, $\bar\beta_1\rightarrow p^+_1$ and $\dl{\bar\beta_1}\rightarrow -iq_1$.}.

Note that for $s=1$ the description in terms of gauge potentials is conformally invariant, even the corresponding action. Furthermore, in this case a manifestly conformal description in terms of potentials can be obtained in our approach by changing the representation for ghosts $\bar\gamma^i$ already at the level of BRST operator \eqref{brst-g}. However, this gives a manifestly conformal description only at the level of equations of motion. The corresponding modif\/ied BRST operator \eqref{brst-g} is not naturally hermitian and hence does not directly determine a manifestly conformal action.

\section{Higher symmetries}

The BRST formulation provides a natural framework to address the
question of describing all symmetries of the equations of motion.
Indeed, for a free f\/ield theory associated to a BRST f\/irst-quantized
system, the BRST state-cohomology at vanishing ghost number corresponds to the {gauge-inequivalent solutions to} the f\/ield equations while the global symmetries can be identif\/ied with the BRST operator-cohomology (i.e.\ the BRST charge acts on operators through the adjoint action). Indeed, the cocycle condition implies that
such an operator preserves the equations of motion, while the
coboundary condition factors out gauge symmetries and on-shell-trivial symmetries.
This is consistent with the f\/irst-quantized point of
view, where global symmetries appear as observables. Strictly speaking,
this identif\/ication applies only to a BRST charge that corresponds to a proper solution of the master equation~(see e.g.~\cite{Henneaux:1992ig} for more details). For instance,
if reducibility relations between constraints are not taken into account
by the BRST charge, then one can in addition have cohomology classes that do not determine nontrivial observables or global symmetries.

The BRST operator cohomology in ghost degree zero is
naturally equipped with an associative product induced by the
operator product of the representatives. The space of
$\mathfrak{osp}(2s|2)$-invariant operators is an associative algebra that
extends the conformal algebra and might prove useful for
constructing interactions for AdS mixed-symmetry gauge f\/ields. In particular, for
$s=0$ this is precisely the well known bosonic higher spin algebra
that underlies the nonlinear theory~\cite{Vasiliev:2003ev} of symmetric f\/ields on AdS space (see also~\cite{Vasiliev:2004cm} for
other higher-spin (super)algebras).

\subsection{Higher symmetries as invariants}

Let us show how the known classif\/ication~\cite{Eastwood:2002su} of
all higher symmetries of the conformal scalar f\/ield can be obtained in our
framework and how this extends to spin-$s$ singletons.
In order to do this, we identify global symmetries with
(a subalgebra of) the operator cohomology for $\brst=\nabla+\bar\brst$ in the space of ghost number zero operators. In turn, this cohomology can be shown to be determined by the cohomology of the f\/iber part $\bar\brst$.

Instead of operators we work with Weyl
symbols so that the operator composition is replaced by the Weyl star
product denoted by $*$. Furthermore, we restrict to operators represented by polynomials in all variables (in particular $Y$) and hence can safely replace $Y+V$ with $Y$
in the expressions for symbols\footnote{Note, however,
that the dependence on $V$ is to be restored once the action on states is concerned.
Indeed, the space of states necessarily contains formal series in $Y$ for which $Y \to Y+V$ is in general ill-def\/ined.}. The adjoint action of $\bar\brst$ is given by
\begin{gather}
  D=\frac{1}{\hbar}\,\qcommut{\,\bar\brst}{\cdot}=
  \frac{1}{\hbar}\,C^I\qcommut{\,t_I}{\cdot}+\text{ghosts} +t_I\dl{\cP_I}+
  \cal{O}(\hbar), \label{eq:37}
\end{gather}
where ``ghosts'' originate from the adjoint action of the term cubic in ghosts of (\ref{barbrst})
{and} the formal quantization parameter $\hbar$ has been introduced
for future convenience. Here $\qcommut{A}{B}=A*B\mp B*A$ is the graded Weyl star-commutator. Recall that $t_I$ denote the symbols of $\mathfrak{osp}(2s|2)$
generators and that $\cP_I$ are ghost momenta associated to~$C^I$.

A natural way to compute the cohomology of the operator $D$ is to compute f\/irst the
cohomology of the Koszul-type dif\/ferential $\delta=t_I\dl{\cP_I}$ by using
homogeneity in $\cP_I$ as a degree and {working} at lowest order in $\hbar$. In
degree $0$, the cocycle condition is trivial while the coboundary
condition can be used to take all representatives totally traceless
with respect to the ambient metric $\eta_{AB}$. Indeed, the Weyl symbols of $t_I$ exhaust all possible
quadratic $\mathfrak{o}(d,2)$-invariants built from the variables
$Y^A$, $\vt^A_i$ and their conjugate momenta (i.e.\ the variables $Z_\alpha^A$ of Subsection~\ref{singl}).
Note that the assumption of using $\hbar$-expansion can be avoided. Indeed in the space of polynomials in~$Z$
one can prescribe $\deg{(Z_\alpha^A)}=-1$ so that $t_I\dl{\cP_I}$ becomes a lowest degree operator to all orders in~$\hbar$
and hence the cohomological problem can again be reduced to the cohomology of~$\delta$.

Let assume for the moment that the cohomology of {$\delta=t_I\dl{\cP_I}$} vanishes
in nonzero degree. Then, the remaining cohomology problem reduces to the one of the dif\/ferential $C^I\qcommut{t_I}{\cdot}+\text{``ghosts''}$, in the space of
$\cP_I$-independent traceless elements. Indeed the last two terms from
\eqref{eq:37} do not contribute. For the reduced problem, ghost number
zero elements are $C^I$-independent because there are no more
variables of negative ghost number left. The coboundary condition is
then trivial while the cocycle condition implies that the cohomology is given by
$\mathfrak{o}(d,2)$-traceless $\mathfrak{osp}(2s|2)$-invariants, i.e., operators commuting with all
generators $t_I$ in this representation and represented by traceless symbols.

In particular, for a conformal scalar f\/ield the cohomology of
$t_I\dl{\cP_I}$ is concentrated in degree~$0$. To see this one observes
that $t_I\dl{\cP_I}$ is dual to the operator from~\cite{Barnich:2004cr}, whose cohomology
was proved in~\cite{Barnich:2004cr} to be concentrated at vanishing ghost number. The global symmetries are then classif\/ied by $\mathfrak{sp}(2)$-invariants in the space of
$\mathfrak{o}(d,2)$-traceless polynomials in the variables $Y^A$ and their conjugate momenta
$P_A$ (i.e.\ the variables $W_m^A$ of Subsection~\ref{cfsc}). These Weyl symbols are represented by traceless rectangular Young tableaux with two rows.
Their characterization and the fact that they exhaust all possible symmetries of a scalar singleton was f\/irst shown in~\cite{Eastwood:2002su} via conformal geometry techniques.

In general, there can be cohomology classes of $t_I\dl{\cP_I}$ in
nonvanishing degree in ghost momenta~$\cP_I$. However, these classes merely ref\/lect that
there are relations between the constraints~$t_I$. In this case
$\bar\Omega$ is not a proper BRST operator and is to be corrected by
terms taking this reducibility into account through appropriate extra
ghost variables. In particular, the Koszul-type dif\/ferential is to be
replaced with the Koszul--Tate dif\/ferential for which the cohomology is indeed
concentrated in degree zero. This shows that possible cohomology
of $t_I\dl{\cP_I}$ in nonvanishing degree is an artifact of using the
nonproper BRST operator and hence should not contribute to global
symmetries.  Of course this argument is not a rigorous mathematical
proof but is a strong indication that $\mathfrak{o}(d,2)$-traceless
$\mathfrak{osp}(2s|2)$-invariants exhaust all possible global
symmetries.

\subsection{Classif\/ication of invariants}

For $s>0$, the $\mathfrak{osp}(2s|2)$-invariants are completely
classif\/ied in the literature~\cite{Sergeev,Sergeev2}. Besides those
generated by invariant bilinears there are some additional invariants.
They do not contribute if, in addition, the f\/ieldstrength is required
to be (anti)-selfdual in each column, as is required for irreducibility.

The
$\mathfrak{osp}(2s|2)$-invariant polynomials in $Z^A_\alpha$
constructed via contraction of all Greek indices through the
symplectic form $\epsilon^{\alpha\beta}$ are polynomials in
the bilinears $J^{AB}=\epsilon^{\alpha\beta}
Z^A_\alpha Z^B_\beta$. Therefore, they are symbols of operators corresponding
to the enveloping algebra of $\mathfrak{o}(d,2)$, i.e.\ polynomials
in the inf\/initesimal generators of the conformal algebra. The
theorem of \cite{Sergeev2,ChengWang} on plethysms (i.e.\ symmetric tensor products) of {rank-two} graded-symmetric
tensors of $\mathfrak{gl}(2s|2)$, such as the metric
$\epsilon^{\alpha\beta}$, gives the structure of the polynomials in~$J^{AB}$. Namely, these {polynomials} decompose into
irreps of $\mathfrak{gl}(d+2)$ labeled by
Young diagrams such that $(i)$~all columns are of even length not greater than
$d+2$, and $(ii)$~any column on the right of the $2s$-th column is
of length two. Moreover, this decomposition is multiplicity-free~\cite{Sergeev2,ChengWang}. Therefore the $\mathfrak{o}(d,2)$-traceless $\mathfrak{osp}(2s|2)$-invariant polynomials correspond to
irreps of $\mathfrak{o}(d,2)$ labeled by the previous Young diagrams
with the supplementary condition that the sum of the lengths of any two columns is not greater than $d+2$. Again, such a decomposition is multiplicity-free.
In the particular cases of spin $s=0$ (and any dimension)~\cite{Eastwood:2002su} or dimension $d=4$ (and any spin) \cite{Vasiliev:2001zy,Anco:2003ze,Pohjanpelto:2008st}, we recover the known facts that all Young diagrams are rectangles made of two rows of equal length.

Actually, there exist $\mathfrak{osp}(2s|2)$-invariant polynomials in the variables $Z^A_\alpha$
which are \textit{not} polynomials in
the bilinears $J^{AB}$. It was shown in \cite{Sergeev,Sergeev2} that $\mathfrak{osp}(2s|2)$-invariants
can be classif\/ied according to the action of the orthogonal group $O(2s)$. The $O(2s)$-action is obtained as follows: the space of all polynomials in $Z^A_\alpha$ is a tensor representation of $\mathfrak{o}(2s)\subset \mathfrak{osp}(2s|2)$ and hence the $\mathfrak{o}(2s)$-action
naturally extends to the representation $\rho$ of the respective group $O(2s)$.
Any $\mathfrak{o}(2s)$-invariant (in particular,
$\mathfrak{osp}(2s|2)$-invariants) is an invariant of $O(2s)$ modulo sign only.
More precisely, the space $\mathfrak{U}$ of all $\mathfrak{osp}(2s|2)$-invariant polynomials in $Z^A_\alpha$ is equipped with
a $\ZZ_2$-grading def\/ined as
\begin{gather*}
\rho(g){\cal O}={\cal O} \qquad \text{for} \quad {\cal O}\in \mathfrak{U}_0,\qquad
\rho(g){\cal O}=\det(g){\cal O} \qquad \text{for} \quad {\cal O}\in \mathfrak{U}_1,
\end{gather*}
for all $g\in O(2s)$.
All invariants in $\mathfrak{U}_0$
are generated by the invariant bilinears $J^{AB}$.
All invariants in $\mathfrak{U}_1$ will be called ``chiral'' because they correspond to the chiral symmetries discussed in~\cite{Pohjanpelto:2008st}, in the sense that they map selfdual states to antiselfdual ones.

Let us now show that if self-duality conditions $*_k\phi=\phi$ (for all $k=1,2,\ldots,s$) are imposed on f\/ields, then
the chiral invariants do not determine symmetries of the restricted equations of motion.
For a f\/ixed column index $k$, let us consider the following transformation $\sigma_k$ from $O(2s)$:
$\sigma_k(\vt_k)=\imath \pi_k$ and
$\sigma_k(\pi_k)=-\imath \vt_k$ while all other variables are left unchanged. Its determinant is equal to $-1$. Working with the dif\/ferential operators instead of their Weyl symbols, the same transformation $\sigma_k$ on operator $\cal O$ can be represented as
\begin{gather*}
 \sigma_k({\cal O})=*_k {\cal O} *_k.
\end{gather*}
Here the duality operator on the $k$-th column is def\/ined as
\begin{gather*}
*_k = \sum\limits_{p=0}^{d+2}\frac{\imath^{p}}{p!\big(d+ 2-p\big)!} \epsilon_{A_1\ldots A_{d+2}} \vt_k^{A_1}\cdots\vt_k^{A_p}
 \dl{\vt_{k\,\,A_{p+1}}}\cdots \dl{\vt_{k\,\,A_{d+2}}},
\end{gather*}
with no sum over the column index $k=1,2,\ldots, s$.
The duality operator $*_k$ can be seen as Hodge conjugation
(modulo a factor $\imath$ when the number of fermionic variables is odd) in the exterior algebra generated by $\vt_k$. Moreover, it squares to the identity $(*_k)^2\phi=\phi$
For an operator~$\cal O$ not involving $\vt_k$ or $\dl{\vt_k}$ the operator $\sigma_k$ acts trivially, because the relation above
amounts to $\sigma_k({\cal O})={\cal O}(*_k)^2={\cal O}$. It is possible to check the action of $\sigma_k$ given above on the basic operators~$\vt_k$ and~$\dl{\vt_k}$.

Let $\phi$ be a solution of the equations of motion and of the selfduality conditions $*_k \phi=\phi$. And let ${\cal O}$ be a chiral symmetry: $\sigma_k({\cal O})=-{\cal O}$ since $\det(\sigma_k)=-1$. One has
\begin{gather*}
*_k ({\cal O}\phi)=(*_k {\cal O} *_k)(*_k \phi)=
\sigma_k({\cal O})\phi=-{\cal O}\phi,
\end{gather*}
where we have used the selfduality condition on $\phi$ and the chirality condition on $\cal O$. This implies that chiral symmetries do not preserve the selfduality condition and hence are not symmetries of the selfdual f\/ields.

Finally, the algebra of all higher symmetries for an irreducible spin-$s$
bosonic singleton should be generated only by the bilinear
$\mathfrak{osp}(2s|2)$-invariants
$J^{AB}=\epsilon^{\alpha\beta}
Z^A_\alpha Z^B_\beta$ modulo the bilinear $\mathfrak{o}(d,2)$-traces $t_{\alpha\beta}=\eta_{AB}Z^A_{\alpha}Z^B_{\beta}$.
This algebra of higher symmetries
is the envelopping algebra of the conformal
algebra $\mathfrak{o}(d,2)$ realized on the singleton module, i.e.\ it is isomorphic to the quotient of ${\cal U}\big(\mathfrak{o}(d,2)\big)$ by the annihilator of the spin-$s$ singleton module.
It is well known that the quotient of the universal enveloping algebra of the spacetime symmetry algebra by the annihilator of the space of solutions is a subalgebra of higher symmetries, but the explicit identif\/ication of the corresponding symmetries as $\mathfrak{o}(d,2)$-traceless $\mathfrak{osp}(2s|2)$-invariants was not known to the best of our knowledge.
Moreover, we provide strong indications that this algebra actually exhaust \textit{all} higher symmetries.
This property generalizes the case $s=0$, i.e.\ the scalar singleton, where it is known that the $\mathfrak{o}(d,2)$-traceless $\mathfrak{sp}(2)$-invariants span all higher symmetries~\cite{Eastwood:2002su}.
For a~\textit{reducible} singleton (i.e.\ if selfduality is not imposed) $\mathfrak{osp}(2s|2)$-invariants contain
so-called chiral symmetries that do not belong to the enveloping of the conformal algebra.

Another generalization which has been achieved in the paper is the identif\/ication, for any integer spin-$s$ singleton,
of non-chiral symmetries (which form a particular enveloping of $\mathfrak{o}(d,2)$ algebra) as an
$\mathfrak{o}(d,2)$-module: the algebra of higher symmetries is a completely reducible $\mathfrak{o}(d,2)$-module which decomposes as the sum of all
irreducible $\mathfrak{o}(d,2)$-modules labeled by Young diagrams such that $(i)$~all columns are of even length, $(ii)$~the sum of the lengths of any two columns is not greater than $d+2$, and $(iii)$~any column on the right of the $2s$-th column is
of length two, where each such irreducible module appears with multiplicity one. This identif\/ication genera\-li\-zes the cases $s=0$ (and $d\geqslant 3$)  \cite{Eastwood:2002su} or dimension $d=4$ (and any spin~$s$) \cite{Vasiliev:2001zy,Anco:2003ze,Pohjanpelto:2008st}, where all Young diagrams are rectangles made of two rows of equal length.

\section*{Acknowledgements}
\label{sec:acknowledgements}

We are grateful to G.~Barnich for collaboration on the initial stage of this project and to Universit\'e Libre de Bruxelles and International Solvay Institutes for their kind hospitality while part of this work was being carried over. We are also grateful to A.~Sergeev for some explanations of his works~\cite{Sergeev,Sergeev2}.
X.B.\ acknowledges M.A.~Vasiliev for helpful comments on his paper~\cite{Vasiliev:2001zy}, M.~Tsulaia for useful discussions, and especially N.~Boulanger for early collaboration on manifestly
conformal wave equations for singletons of integer spin. M.G.\
thanks G.~Bonelli, E.~Feigin, R.~Metsaev and M.A.~Vasiliev for discussions.
The work of M.G.\ is supported by the RFBR grant
10-01-00408 and RFBR-CNRS grant 09-01-93105.


\pdfbookmark[1]{References}{ref}
\LastPageEnding

\end{document}